\documentclass[prb,amsmath,groupedaddress,showpacs,showkeys,twocolumn]{revtex4-1}
\usepackage{amssymb,amsmath}   % need for subequations
\usepackage[dvips]{graphicx}   % need for figures
\usepackage{verbatim}   % useful for program listings
\usepackage{color}      % use if color is used in text
\usepackage{subfigure}  % use for side-by-side figures
\usepackage{hyperref}   % use for hypertext links, including those to external documents and URLs
\usepackage{gensymb}
\usepackage{epstopdf}
\usepackage{natbib}
\usepackage{enumerate}
\usepackage{braket}

\begin{document}

\title{Bond-counting potentials --- A classical many-body model of covalent bonding with exact solutions in one dimension}

\author{Shlomi Matityahu}
\email{matityas@post.bgu.ac.il} 
\affiliation{Department of Physics, NRCN, P.O. Box 9001, Beer-Sheva 84190, Israel}
\author{Nathan Argaman}
\email{argaman@mailaps.org}
\affiliation{Department of Physics, NRCN, P.O. Box 9001, Beer-Sheva 84190, Israel}

\date{\today}
\begin{abstract}
We introduce ``bond-counting'' potentials, which provide an elementary description of covalent bonding. These simplistic potentials are intended for studies of the mechanisms behind a variety of phase transitions in elemental melts, including the liquid-liquid phase transitions (LLPT) in phosphorus and bismuth. As a first study employing such potentials, an analytic solution of a one-dimensional model system is presented, including its thermodynamic properties and its structure factor. In the simplest case, the chemical valency of each atom is $1$, and either single atoms or diatomic molecules are present. At low temperatures and moderate pressures, the system consists almost exclusively of molecules, and single atoms act as topological defects. A slightly more complicated case involves a valency of $2$, with either single or double bonding. This system exhibits a first-order LLPT from a molecular to a polymeric phase, as in phosphorus. In this case, the one-dimensional model system exhibits phase separation for finite-sized systems at low temperatures. A variant of this system also exhibits a non-equilibrium phase transformation upon heating the molecular condensed phase, qualitatively similar to boiling in white phosphorus.

\end{abstract}

%\pacs{85.75.Hh, 75.76.+j, 72.25.Dc, 75.70.Tj}
 \maketitle
\section{Introduction} \label{Introduction}

Many liquids and solids are well-described within the van der Waals picture: their structure and dynamics are determined primarily by strong repulsive forces that give rise to severe constraints on the particles' motion, whereas the overall density is largely determined by weaker long-range attractive forces.~\cite{Hansen&McDonald,CD83} The packing constraints dominate the short-range order characteristic of the liquid state, and lead to a quasi-universal structure of simple liquid metals and noble gases, at least in the vicinity of the melting curve, with a structure close to that of a system of hard spheres, and coordination numbers of $z=9-11$.~\cite{Hansen&McDonald,ANW66,YJL73,DJC16}

On the other hand, covalent-bonded elements with open crystalline structures --- mainly group IV, V, and VI elements --- exhibit liquid structures characterized by low coordination numbers. The fact that such a liquid structure may be highly non-trivial is reflected by intricate phase diagrams, which include thermodynamic anomalies and/or liquid-liquid phase transitions (LLPTs).~\cite{BVV97,BVV02,BVV03,MPF07} Indeed, a growing number of LLPTs have been observed in elemental liquids during the last three decades, e.g., in phosphorus,~\cite{KY00,KY02,MG03} selenium,~\cite{BVV89,BVV91} sulfur,~\cite{BVV91,LL14,ZG14,HL18}, bismuth~\cite{UAG92,GY09,SY17,EM18} and tellurium.~\cite{TY85,BVV92} 

In some cases, there exists a general understanding of the mechanism behind an LLPT. The most well-known example is phosphorus, which at temperatures of about $1300\,$K and moderate pressures exhibits a fluid state consisting of pyramid-shaped $P_4$ molecules. As the pressure is increased to about $1\,$GPa, the volume occupied by the pyramidal molecules becomes prohibitive, and the liquid collapses to a much denser polymeric structure.~\cite{KY00,KY02} This LLPT is first-order and reversible, with a coexistence line ranging from $(P,T)\approx(0.9\,\mathrm{GPa},1300\,\mathrm{K})$ to $(P,T)\approx(0.3\,\mathrm{GPa},2450\,\mathrm{K})$.~\cite{MG03}

A further consequence of strong covalent bonds is the observation of non-equilibrium phase transitions between metastable states.~\cite{BVV06} A prominent example of an element with such states is, again, phosphorus.~\cite{Phase Diagrams} The $P_4$ molecules can be condensed from the vapor, and the resulting material is known as white phosphorus. White phosphorus exists in several different molecular phases as a function of temperature, and undergoes solid-solid, melting and boiling transitions between these metastable phases, despite the fact that the thermodynamically stable phase under these conditions is the non-molecular black phosphorus state. In the vicinity of room temperature, the lifetimes of the metastable phases are exceedingly long, but near boiling, at $\approx 550\,\mathrm{K}$, polymerization of phosphorus begins to occur,~\cite{DTW46} and the original $P_4$ molecules do not reform upon subsequent cooling to room temperature conditions. The present description of the phase behavior of phosphorus does not do justice to the complexity of the topic (e.g., we have not even mentioned red phosphorus). Instead, it focuses exclusively on those aspects --- the LLPT and the existence of a non-equilibrium boiling transformation --- which, as we will see below, can be qualitatively reproduced within a simplistic model for covalent bonding in one dimension (1D).

In other cases, the mechanisms behind observed LLPTs have not yet been identified. For example, different types of evidence for several LLPTs in bismuth have been reported, both at high pressures~\cite{UAG92,EM18} and at ambient pressure.~\cite{LXF07,GY09} A recent set of experiments presents further challenges for an understanding of the behavior of this elemental liquid.\cite{SY17,SY16a,SY16b} In these studies, bismuth is compressed to $\sim 2\,$GPa and heated to $\sim 2000\,$K, and then cooled and decompressed back to ambient conditions. The resulting solid contains structural defects and displays a variety of anomalies, including the following peculiarity at melting: upon heating at ambient pressure, this material undergoes a transformation similar to standard melting of Bi I (as evidenced by an endothermic peak which has the appropriate magnitude and temperature, and is largely reversible), but the material does not flow and instead retains its shape like a solid [see the insets in Fig.~1 of Ref.~\onlinecite{SY16b}]. Identifying the mechanism behind such a phenomenon is a challenge which requires significant experimental and theoretical work. We will present some speculative ideas on this in Sec.~\ref{Discussion}, and explain in what sense the present work may serve as a preliminary step in a 3D-simulation approach to this problem.

Several theoretical approaches have been used to investigate LLPTs. On the phenomenological level, the two-state model~\cite{SS65,RE67} was used to give a simple explanation of LLPTs, for both liquids and amorphous solids.~\cite{BVV97,TH00,PEG03} According to this model, an LLPT is a result of competition between two kinds of clusters which differ in their short-range order. One liquid phase consists predominantly of entropically favorable clusters of high density, and the second phase consists of a large concentration of energetically favorable clusters of low density (i.e., with a more open structure).~\cite{twostatevariants} The model assumes an energy cost $J$ for mixing the two types of clusters, and yields a first-order phase transition line, terminating at a critical point at a temperature of $J/(2k^{}_{\mathrm{B}})$. At lower temperatures, the two types of clusters exhibit phase separation despite being made of the same substance (i.e., exhibit a miscibility gap). This type of modeling can describe the thermodynamics of the system without requiring a microscopic understanding of the mechanisms of formation and interaction of the two distinct clusters involved.

More elaborate theoretical analyses employ molecular dynamics simulations, either combined with ab-initio density functional theory calculations~\cite{MT01,GM05} or based on empirical potentials.~\cite{PHP03,SS03} Most of these simulations strive to achieve a quantitative understanding of the phenomena in specific materials, and are often successful, but are limited by computational power for the first type of simulation, and by the accuracy of the potentials for the latter type. For example, the predictions of LLPTs by such simulations sometimes contradict experimental observations [see, e.g., the case of nitrogen in Ref.~\onlinecite{WG17}], indicating possible gaps in our understanding. A complementary approach in seeking a better understanding of LLPTs, is based on simplistic potentials with very few parameters, where the goal is to reproduce observed phenomena qualitatively, rather than quantitatively.

The canonical example of such a simplistic approach to the liquid state of matter is the hard sphere potential, with only one parameter (the spheres' diameter), which provides an elementary prototype of the melting transition.~\cite{ABJ57,WWW57,HWG68} More recently, a variety of simple soft-core isotropic pair potentials have been investigated extensively, and have been shown to exhibit a wide range of non-trivial phenomena, including LLPTs, polyamorphism in glasses, anomalous melting and water-like liquid anomalies [see Ref.~\onlinecite{BSV09} and references therein]. In some cases, such as the density maximum in water, the phenomenon of interest could be qualitatively reproduced even within a simplified 1D model, by analytic solution of the thermodynamics of the system.~\cite{CCH96,SLMR99,BNA08}  

In the present work we introduce ``bond-counting'' many-body potentials, in order to provide a simple description of covalent bonding. In the simple cases studied in detail, each atom is characterized by two size parameters --- a ``core diameter'' which is impenetrable, and a larger ``bonding zone diameter'' which may be penetrated by up to $z$ neighbors, each forming a bond with a binding energy of one unit. The case $z=1$ corresponds to atoms which form diatomic molecules, such as hydrogen or fluorine; $z=2$ describes atoms which can polymerize, such as sulfur, and $z=3$ may describe bismuth and phosphorous (in 3D). The present contribution is limited to 1D, where analytical solutions are possible. The thermodynamics and structure of the simplest model with $z=1$ are analyzed in detail. Subsequently, a qualitative description of the above-mentioned phenomena observed in phosphorus is obtained for $z=2$. This is done by allowing for double bonds between identical atoms, which opens the possibility of competing molecular and polymeric configurations even in a very simple 1D system. 

The paper is organized as follows. Sec.~\ref{Model} provides the general definition of potentials which count the number of bonds. The exact solution of this model in 1D is discussed in Sec.~\ref{Solution}. In Sec.~\ref{Discussion} we summarize the results and discuss possible extensions of this work, including higher dimensions.

\section{Bond-counting model potentials for covalent bonding} \label{Model}

Before providing the definition of a new family of potentials, it is appropriate to mention a few more general facts regarding existing empirical potentials.~\cite{Hansen&McDonald} First, it is noteworthy that both qualitative and remarkable quantitative accuracy in the description of simple liquids, consisting of atoms with van der Waals, metallic or ionic interactions can be achieved using simple pair potentials, such as the Lennard-Jones or (possibly screened) Coulomb potentials. For molecular liquids, models which combine pair potentials between atoms in different molecules with rigid connections within each molecule are also successful, as long as breaking and reformation of the covalent bonds is not relevant. Of the very many multi-parameter many-body empirical potentials developed to improve the quantitative accuracy of simulations, we will mention only two: the embedded-atom method~\cite{DMS93} (EAM), to which we will return below, and reactive force fields,~\cite{STP16} which were developed expressly in order to allow simulations of covalent bond breaking and formation.

Liquids of the latter type, for which the bonds within each molecule are evolving (as in polymerization), are referred to as ``associating liquids''. A simple model for these describes each atom or monomer as a hard sphere, augmented by several much smaller off-center spheres with attractive square-well potentials.~\cite{WMS84,JG88} If the attractive potential is sufficiently short-ranged, the formation of more than one bond at each off-center site is excluded. Approximations appropriate for such models have been developed,~\cite{WMS84} and have seen considerable success in the framework of the statistical associating fluid theory (SAFT).~\cite{CWG} The bond-counting potentials introduced here express the idea of chemical valency directly, rather than relying on geometric constraints and pair potentials. In particular, the extension to double or triple bonds is straightforward within the bond-counting approach, see below.

In order to define bond-counting potentials, consider a system of $N$ classical particles ("atoms") of mass $m$ and positions $\mathbf{r}^{}_{i}$. Distinguishing distances shorter than the bond-range $a$ from longer distances, each atom is associated not only with its position, but also with the number of its bonds 
\begin{align}
\label{eq:Number_of_bonds1}
&q^{}_{i}=\sum^{}_{k\neq i}\theta(a-r^{}_{ik}),
\end{align}
where $r^{}_{ik}=|\mathbf{r}^{}_{i}-\mathbf{r}^{}_{k}|$ is the distance between atoms $i$ and $k$ and $\theta(x)$ is the step function. The valency $z$ of each atom cannot be exceeded, namely $q^{}_{i}\leq z$, making this a many-body potential. In addition, consider a potential $\mathcal{V}^{}_{ij}$ between each pair of atoms $i$ and $j$ which may, in general, depend on the number of bonds of both, namely $\mathcal{V}^{}_{ij}=\mathcal{V}(r^{}_{ij},q^{}_{i},q^{}_{j})$. In the simple cases to be studied here, the potential is taken to be independent of $q^{}_{i}$ and $q^{}_{j}$, provided that both are not larger than $z$:
\begin{align}
\label{eq:Bond_counting_potential1}
&\mathcal{V}(r^{}_{ij},q^{}_{i},q^{}_{j})=\begin{cases}
v(r^{}_{ij}) & q^{}_{i},q^{}_{j}\leq z \\ 
\infty & q^{}_{i}>z\;\,\mathrm{or}\;\, q^{}_{j}>z
\end{cases},
\end{align}
where
\begin{align}
\label{eq:Bond_counting_potential2}
&v(r)=\begin{cases}
\infty & r<d \\ 
v^{}_{\mathrm{bond}}(r) & d<r<a \\
v^{}_{\mathrm{non-bond}}(r) & r>a
\end{cases}.
\end{align}
Here, $d$ is an impenetrable core diameter (which is optional in the sense that $d=0$ is allowed), and $v^{}_{\mathrm{bond}}(r)$, $v^{}_{\mathrm{non-bond}}(r)$ are the interaction potentials inside and outside the bonding zone, respectively.

Double bonds can be taken into account in a straightforward manner within this approach, by introducing an additional diameter $\tilde{a}<a$, such that the region $d<r<\tilde{a}$ corresponds to a double bond. The number of bonds of atom $i$ is then 
\begin{align}
\label{eq:Number_of_bonds2}
&q^{}_{i}=\sum^{}_{k\neq i}\left[\theta(a-r^{}_{ik})+\theta(\tilde{a}-r^{}_{ik})\right],
\end{align}
and the potential $v(r)$ in Eq.~(\ref{eq:Bond_counting_potential1}) is replaced by
\begin{align}
\label{eq:Bond_counting_potential3}
&v(r)=\begin{cases}
\infty & r<d \\ 
v^{}_{\mathrm{double-bond}}(r) & d<r<\tilde{a} \\
v^{}_{\mathrm{single-bond}}(r) & \tilde{a}<r<a \\
v^{}_{\mathrm{non-bond}}(r) & r>a
\end{cases}.
\end{align}
Similarly, triple bonds can be described by introducing yet another diameter.

We note that just as hard spheres can be viewed as a limiting form of pair potentials, the bond-counting potentials can be represented as limiting forms of more elaborate many-body potentials. For example, using the EAM with the notation
\begin{align}
\label{eq:Embedded_Atom_Method1}
\mathcal{V}^{}_{\mathrm{EAM}}(\mathbf{r}^{}_{1}, & \ldots,\mathbf{r}^{}_{N})=
\nonumber\\ &
\sum^{N}_{i=1}F\left(\sum_{j\neq i}\rho(r^{}_{ij})\right)+\sum^{N}_{i=2}\sum^{i-1}_{j=1}U(r^{}_{ij}),
\end{align}
one obtains the above potentials by using the choice $\rho(r^{}_{ij})=\theta(a-r^{}_{ij})$ for a model with single bonds ($\rho(r^{}_{ij})=\theta(a-r^{}_{ij})+\theta(\tilde{a}-r^{}_{ij})$ for a model with double bonds), with the embedding energy $F$ and pair potential $U$ given by
\begin{align}
\label{eq:Embedded_Atom_Method2}
&F(x)=\begin{cases}
0 & x\leq z \\ 
\infty & x>z
\end{cases},\nonumber\\
&U(r)=v(r).
\end{align}
Thus, simulation software with an EAM option can be used to obtain results for the present family of potentials, provided that numerical issues associated with the use of step functions are resolved. Note that this limiting case is unconventional in the sense that the embedding energy is purely repulsive, with all the attraction in the pair potential. This is the opposite of the customary use of the EAM, which was originally developed for metals.

It should be emphasized that the family of potentials defined above is isotropic --- there is no explicit dependence on bond angles. As appropriate for liquids, it is suited for identifying mechanisms which rely exclusively on the notion of chemical valency, and are not sensitive to the details of the geometry. Adding angular dependencies, as in the development of the modified EAM,~\cite{Baskes92} is a natural possibility for future studies.

\section{Exact solution in 1D} \label{Solution}

In this section we briefly review the Takahashi solution for a classical system with nearest-neighbor interactions in 1D (Sec.~\ref{Takahashi Solution}), and then use the transfer matrix method to generalize and apply the solution to two models of bond-counting potentials: (1) a model with $z=1$, for which we study in detail the equilibrium statistical mechanics and the resulting structure (Sec.~\ref{z=1}); (2) a model with $z=2$ which allows for double bonds, exhibiting a polymerization transition (Sec.~\ref{z=2}).

\subsection{The Takahashi solution for 1D systems with nearest-neighbor interactions} \label{Takahashi Solution}

We summarize the Takahashi solution~\cite{TH42,Lieb&Mattis} for a 1D classical system of $N$ particles of mass $m$, with Hamiltonian of the form
\begin{align}
\label{eq:Hamiltonian1}
&\mathcal{H}(\mathbf{x},\mathbf{p})=\frac{1}{2m}\sum^{N}_{i=1}p^{2}_{i}+\sum^{}_{\braket{i,j}}u(|x^{}_{i}-x^{}_{j}|).
\end{align}
Here, $x^{}_{i}$ and $p^{}_{i}$ are the position and momentum of the $i$th particle, and $u(x)$ is the nearest-neighbor interaction potential (the sum in the last term is over nearest neighbors $i$, $j$).~\cite{Comment1} 

The canonical partition function for a system of length $L$ at temperature $T$ reads
\begin{align}
\label{eq:Partition_LTN1}
Z^{}_{N}(L,T) & =\frac{1}{N!h^{N}}\int d^{N}xd^{N}p\,e^{-\beta\mathcal{H}(\mathbf{x},\mathbf{p})}\nonumber\\
& = \frac{1}{N!\lambda^{N}_{T}}\int d^{N}x\,e^{-\beta\sum^{}_{\braket{i,j}}u(|x^{}_{i}-x^{}_{j}|)},
\end{align}
where $\beta=1/k^{}_{\mathrm{B}}T$ and $\lambda^{}_{T}=h/\sqrt{2\pi mk^{}_{\mathrm{B}}T}$, with $k^{}_{\mathrm{B}}$ and $h$ the Boltzmann and Planck constants. The crucial step in deriving the exact solution is that in 1D the last integral in~(\ref{eq:Partition_LTN1}) is $N!$ times the integral over the domain $\mathcal{D}$, defined by $0\leq x^{}_{1}\leq x^{}_{2}\leq\ldots\leq x^{}_{N}\leq L$. The canonical partition function then takes the form 
\begin{align}
\label{eq:Partition_LTN2}
&Z^{}_{N}(L,T)=\frac{1}{\lambda^{N}_{T}}\int^{}_{\mathcal{D}}d^{N}x\,e^{-\beta\sum^{N-1}_{i=1}u(x^{}_{i+1}-x^{}_{i})}.
\end{align}

The evaluation of the partition function is simple in the isobaric-isothermal ($PTN$) ensemble, for which 
\begin{align}
\label{eq:Partition_PTN1}
&Z^{}_{N}(P,T)=\frac{1}{l_0}\int^{\infty}_{0}dL\,Z^{}_{N}(L,T)e^{-\beta PL},
\end{align}
where $l_0$ is a basic unit of length required to render the partition function $Z^{}_{N}(P,T)$ dimensionless. In the thermodynamic limit the choice of $l_0$ is not important; below $l_0=\left(\beta P\right)^{-1}$ is used, so that the condition $Z^{}_{N=0}(L,T)=1$ implies $Z^{}_{N=0}(P,T)=1$. Substituting Eq.~(\ref{eq:Partition_LTN2}) into~(\ref{eq:Partition_PTN1}) and changing variables to $y^{}_{1}=x^{}_{1}$, $y^{}_{i}=x^{}_{i}-x^{}_{i-1}$ for $i=2,\ldots,N$ and $y^{}_{N+1}=L-x^{}_{N}$ (such that $\sum^{N+1}_{i=1}y^{}_{i}=L$), one obtains
\begin{align}
\label{eq:Partition_PTN2}
Z^{}_{N}(P,T)=\frac{1}{l_0 \lambda^{N}_{T}}\int^{\infty}_{0}dy^{}_{1}e & ^{-\beta Py^{}_{1}}\times\nonumber\\
\left(\int^{\infty}_{0}dye^{-\beta\left[u(y)+Py\right]}\right)^{N-1}&\int^{\infty}_{0}dy^{}_{N+1}e^{-\beta Py^{}_{N+1}}.
\end{align}
All thermodynamic properties can then be calculated exactly from the Gibbs free energy per particle (or chemical potential), which in the thermodynamic limit is
\begin{align}
\label{eq:Gibbs}
g(P,T)&=\lim^{}_{N\rightarrow\infty}\frac{G^{}_{N}(P,T)}{N}=-\frac{1}{\beta}\lim^{}_{N\rightarrow\infty}\frac{\ln Z^{}_{N}(P,T)}{N}\nonumber\\
&=-\frac{1}{\beta}\ln\left(\frac{1}{\lambda^{}_{T}}\int^{\infty}_{0}dye^{-\beta\left[u(y)+Py\right]}\right).
\end{align}

An example, to be used below as a reference, is the hard-sphere model
\begin{align}
\label{eq:Hard_sphere_potential1}
&u^{}_{\mathrm{HS}}(x)=\begin{cases}
\infty & x < a \\ 
0 & x > a
\end{cases},
\end{align}
for which Eq.~(\ref{eq:Partition_PTN2}) yields
\begin{align}
\label{eq:Partition_PTN_hard_sphere}
&Z^{(\mathrm{HS})}_{N}(P,T)=\frac{1}{\beta P\lambda^{}_{T}}\left(\frac{e^{-\beta Pa}}{\beta P\lambda^{}_{T}}\right)^{N-1}.
\end{align}
The corresponding Gibbs free energy per particle [Eq.~(\ref{eq:Gibbs})] is
\begin{align}
\label{eq:Gibbs_hard_sphere}
g^{}_{\mathrm{HS}}(P,T)=Pa+\frac{1}{\beta}\ln\left(\beta P\lambda^{}_{T}\right).
\end{align}

\subsection{Bond-counting potential with $\mathbf{z=1}$} \label{z=1}

Using the transfer matrix method, the Takahashi solution may be generalized to solve for a system with a $z=1$ bond-counting potential in 1D. The relevant potentials for $z=1$ are given by Eqs.~(\ref{eq:Number_of_bonds1})-(\ref{eq:Bond_counting_potential2}). Each atom can have at most a single bond with one of its nearest neighbors, and the system consists of a mixture of single atoms and diatomic molecules (i.e., paired atoms separated by a distance $d<y<a$). We note that a similar method was used in Ref.~\onlinecite{BNA92} to solve a 1D model involving two types of atoms (or rather, two orientations of molecules). In the solution below the index of the transfer matrix specifies the type of bond, rather than the type of atom as in Ref.~\onlinecite{BNA92}.

\subsubsection{Thermodynamics}

In the $PTN$ ensemble each molecular bond will contribute to the partition function a factor 
\begin{align}
\label{eq:A}
&A=\int^{a}_{d}dye^{-\beta\left[v^{}_{\mathrm{bond}}(y)+Py\right]},
\end{align}
whereas each pair of adjacent atoms which are not bound to each other will contribute a factor
\begin{align}
\label{eq:B}
&B=\int^{\infty}_{a}dye^{-\beta\left[v^{}_{\mathrm{non-bond}}(y)+Py\right]}.
\end{align}
The partition function for $N$ atoms is then $l^{}_{0}\lambda^{-N}_{T}$ times a sum of all possible products of $N-1$ factors of this type (such as $ABABBA$ for $N=7$), where the only restriction on the products is that two consecutive bond factors $A$ cannot occur. An appropriate recursion relation is available, $Z^{}_{N}= \lambda^{-1}_{T}B\,Z^{}_{N-1}+\lambda^{-2}_{T}AB\,Z^{}_{N-2}$, expressing the fact that the last atom in the chain can be either unbound or bonded to the preceding atom: one can add to shorter chains either a single atom or a molecule. This relation is similar to the recursion relation for the Fibonacci sequence, providing a hint that expressions in terms of powers of a $ 2\times 2$ matrix may be possible.

The calculation of the partition function can be simplified by closing the 1D chain, such that the $N$th atom becomes a neighbor of the first atom. One may define the bond variables $\{\sigma^{}_{i}\}^{N}_{i=1}$, where $\sigma^{}_{i}=1$ if atoms $i$ and $i+1$ are bonded (i.e., $d<y^{}_{i+1}<a$) and $\sigma^{}_{i}=0$ otherwise. With this definition, the contribution of the $i$th bond to the partition function is $B$ if $\sigma^{}_{i}=0$, and $A$ if $\sigma^{}_{i}=1$. The restriction that two adjacent bonds are not allowed implies that the contribution vanishes if $\sigma^{}_{i}=\sigma^{}_{i+1}=1$ for some value of $i$. The partition function can thus be written in the form
\begin{align}
\label{eq:Partition_PTN3}
&Z^{}_{N}(P,T)=\frac{1}{\lambda^{N}_{T}}\sum^{}_{\{\sigma^{}_{i}\}}\prod^{N}_{i=1}T^{}_{\sigma^{}_{i}\sigma^{}_{i+1}}=\frac{1}{\lambda^{N}_{T}}\mathrm{Tr}\left[T^{N}\right],
\end{align}
where $\sigma^{}_{N+1}\equiv\sigma^{}_{1}$ and the transfer matrix $T$ is
\begin{align}
\label{eq:Transfer_matrix1}
&T=\begin{pmatrix}
B & B \\
A & 0
\end{pmatrix}.
\end{align}

The eigenvalues of this transfer matrix are $\Lambda^{}_{\pm}=\left(B\pm\sqrt{B^{2}+4AB}\right)/2$ and since $\Lambda^{}_{+}>\Lambda^{}_{-}$, in the thermodynamic limit the Gibbs free energy per particle reads
\begin{align}
\label{eq:Gibbs_bond}
g(P,T)&=-\frac{1}{\beta}\ln\left(\frac{\Lambda^{}_{+}}{\lambda^{}_{T}}\right)\nonumber\\
&=g^{}_{\mathrm{Tak}}(P,T)-\frac{1}{\beta}\ln\left[\frac{1}{2}\left(1+\sqrt{1+\frac{4A}{B}}\right)\right],
\end{align}
where $g^{}_{\mathrm{Tak}}(P,T)=-k^{}_{\mathrm{B}}T\ln\left(B/\lambda^{}_{T}\right)$ is the Gibbs free energy of the Takahashi model [Eq.~(\ref{eq:Gibbs})] with an interaction potential $u(y)$ between a paired and a single nearest-neighbor atoms, 
\begin{align}
\label{eq:nearest_neighbor_potential}
&u(y)=\begin{cases}
\infty & y < a \\ 
v^{}_{\mathrm{non-bond}}(y) & y > a
\end{cases}.
\end{align}
The length $\ell(P,T)$ and entropy $s(P,T)$ per particle are therefore
\begin{align}
\label{eq:Length1}
&\ell=\left(\frac{\partial g}{\partial P}\right)^{}_{T}=\ell^{}_{\mathrm{Tak}}-\frac{\frac{2}{\beta}\frac{\partial}{\partial P}\left(\frac{A}{B}\right)^{}_{T}}{\left(1+\sqrt{1+\frac{4A}{B}}\right)\sqrt{1+\frac{4A}{B}}},\\
\label{eq:Entropy1}
&s=-\left(\frac{\partial g}{\partial T}\right)^{}_{P}=s^{}_{\mathrm{Tak}}+k^{}_{\mathrm{B}}\ln\left[\frac{1}{2}\left(1+\sqrt{1+\frac{4A}{B}}\right)\right]+\nonumber\\
&\qquad\qquad\qquad\qquad\frac{\frac{2}{\beta}\frac{\partial}{\partial T}\left(\frac{A}{B}\right)^{}_{P}}{\left(1+\sqrt{1+\frac{4A}{B}}\right)\sqrt{1+\frac{4A}{B}}},
\end{align}
where $\ell^{}_{\mathrm{Tak}}=\left(\partial g^{}_{\mathrm{Tak}}/\partial P\right)^{}_{T}$ and $s^{}_{\mathrm{Tak}}=-\left(\partial g^{}_{\mathrm{Tak}}/\partial T\right)^{}_{P}$.

As a simple example, let us consider the case $v^{}_{\mathrm{non-bond}}(x)=0$ and $v^{}_{\mathrm{bond}}(x)=-\varepsilon$, where $\varepsilon>0$ is a constant energy. Then $u(y)$ [Eq.~(\ref{eq:nearest_neighbor_potential})] reduces to the hard-sphere potential $u^{}_{\mathrm{HS}}(y)$ [Eq.~(\ref{eq:Hard_sphere_potential1})], and thus $g^{}_{\mathrm{Tak}}=g^{}_{\mathrm{HS}}$. From Eq.~(\ref{eq:Gibbs_hard_sphere}) we obtain
\begin{align}
\label{eq:Length2}
&\ell^{}_{\mathrm{Tak}}=\ell^{}_{\mathrm{HS}}=a+\frac{1}{\beta P},\\
\label{eq:Entropy2}
&s^{}_{\mathrm{Tak}}=s^{}_{\mathrm{HS}}=k^{}_{\mathrm{B}}\left[\frac{3}{2}-\ln\left(\beta P\lambda^{}_{T}\right)\right].
\end{align}
Note that Eq.~(\ref{eq:Entropy2}) is also the entropy per particle of a 1D ideal gas at fixed pressure $P$ and temperature $T$, and that the hard-sphere diameter in Eq.~(\ref{eq:Length2}) is the bond range $a$ rather than the core diameter $d$.

The ratio $A/B$ for this system is
\begin{align}
\label{eq:A_over_B}
&\frac{A}{B}=e^{\beta\epsilon}\left[e^{\beta P\left(a-d\right)}-1\right].
\end{align}
In the high density (i.e., low temperature or high pressure) limit $\beta P\left(a-d\right)\gg 1$ Eqs.~(\ref{eq:Length1})-(\ref{eq:A_over_B}) give 
\begin{align}
\label{eq:High_density_limit}
&\ell=\frac{a+d}{2}+\frac{1}{\beta P}+(a-d)\,\mathcal{O}(e^{-\beta P\left(a-d\right)}),\nonumber\\
&s=s^{}_{\mathrm{HS}}+k^{}_{\mathrm{B}}\,\mathcal{O}(e^{-\beta P\left(a-d\right)}).
\end{align}
This limit corresponds to a close-packed arrangement of $N/2$ molecules, each of diameter $a+d$. The system then behaves effectively as $N$ hard spheres each of diameter $\left(a+d\right)/2$. In the low density and low temperature limit, $e^{-\beta\varepsilon}\ll\beta P\left(a-d\right)\ll 1$, we obtain
\begin{align}
\label{eq:low_Temperature_density_limit}
2\ell=\frac{3a+d}{2}+\frac{1}{\beta P}+(a-d)\,\mathcal{O}(\beta P\left(a-d\right)).
\end{align}
In this limit the length of each molecule follows the equation of state~(\ref{eq:Length2}) of hard spheres with diameter $\left(3a+d\right)/2$, corresponding to the average diameter of each molecule. In the low density and high temperature limit, $\beta P(a-d)\ll e^{-\beta\varepsilon}$, we find
\begin{align}
\label{eq:High_Temperature_limit}
\ell=&\,a\left(1-e^{\beta\varepsilon}\right)+de^{\beta\varepsilon}+\frac{1}{\beta P}+\nonumber\\
&(a-d)\,\mathcal{O}(\beta P\left(a-d\right)e^{\beta\varepsilon}),\nonumber\\
s=&\,s^{}_{\mathrm{HS}}+k^{}_{\mathrm{B}}\,\mathcal{O}(\beta P\left(a-d\right)e^{\beta\varepsilon}),
\end{align}
which, if the temperature is also high in the sense that $\beta\varepsilon\ll 1$, corresponds to a system of $N$ hard spheres of diameter $d$.

\begin{figure}[ht]
	\centering
	\includegraphics[width=0.52\textwidth,height=0.28\textheight]{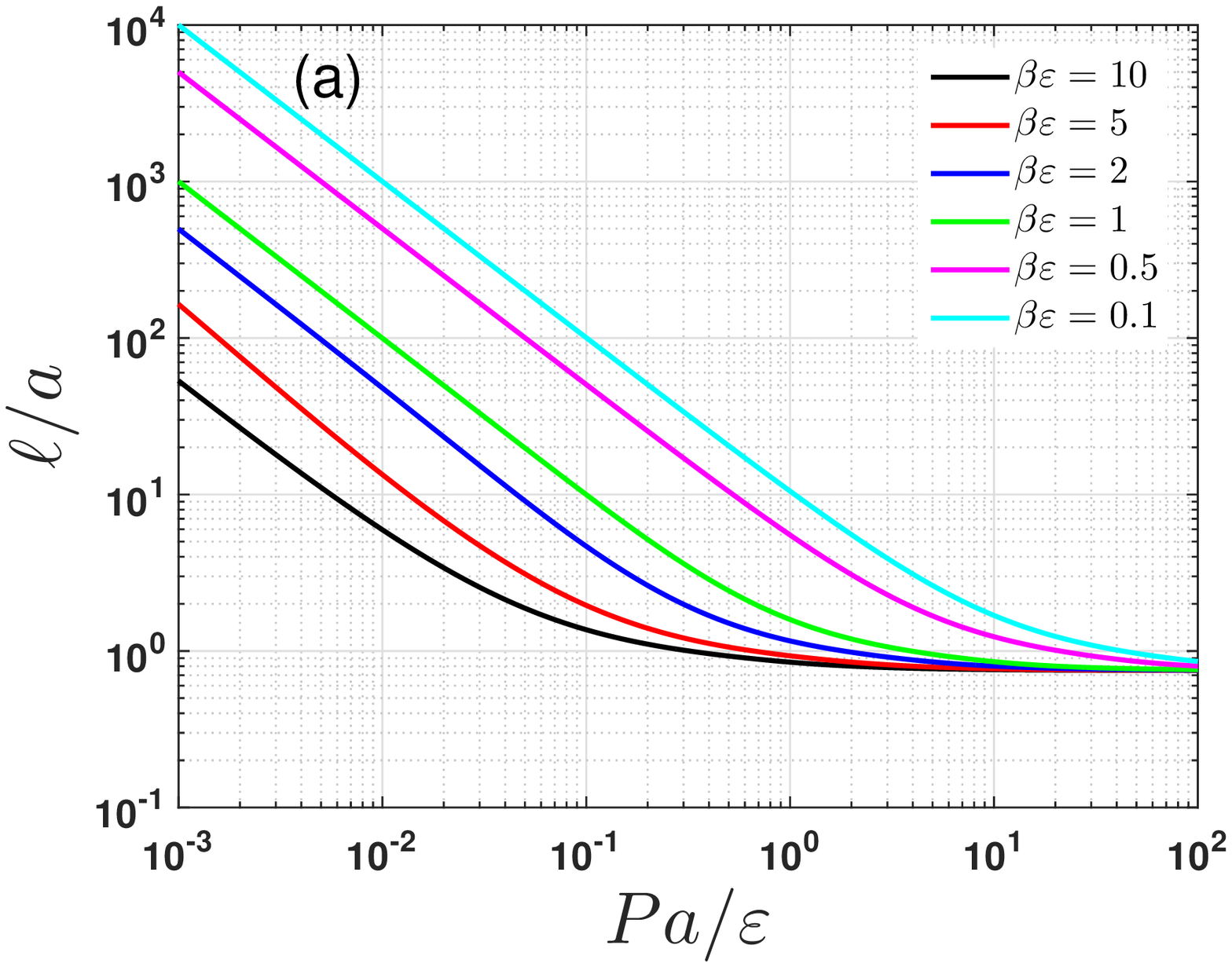}
	\includegraphics[width=0.52\textwidth,height=0.28\textheight]{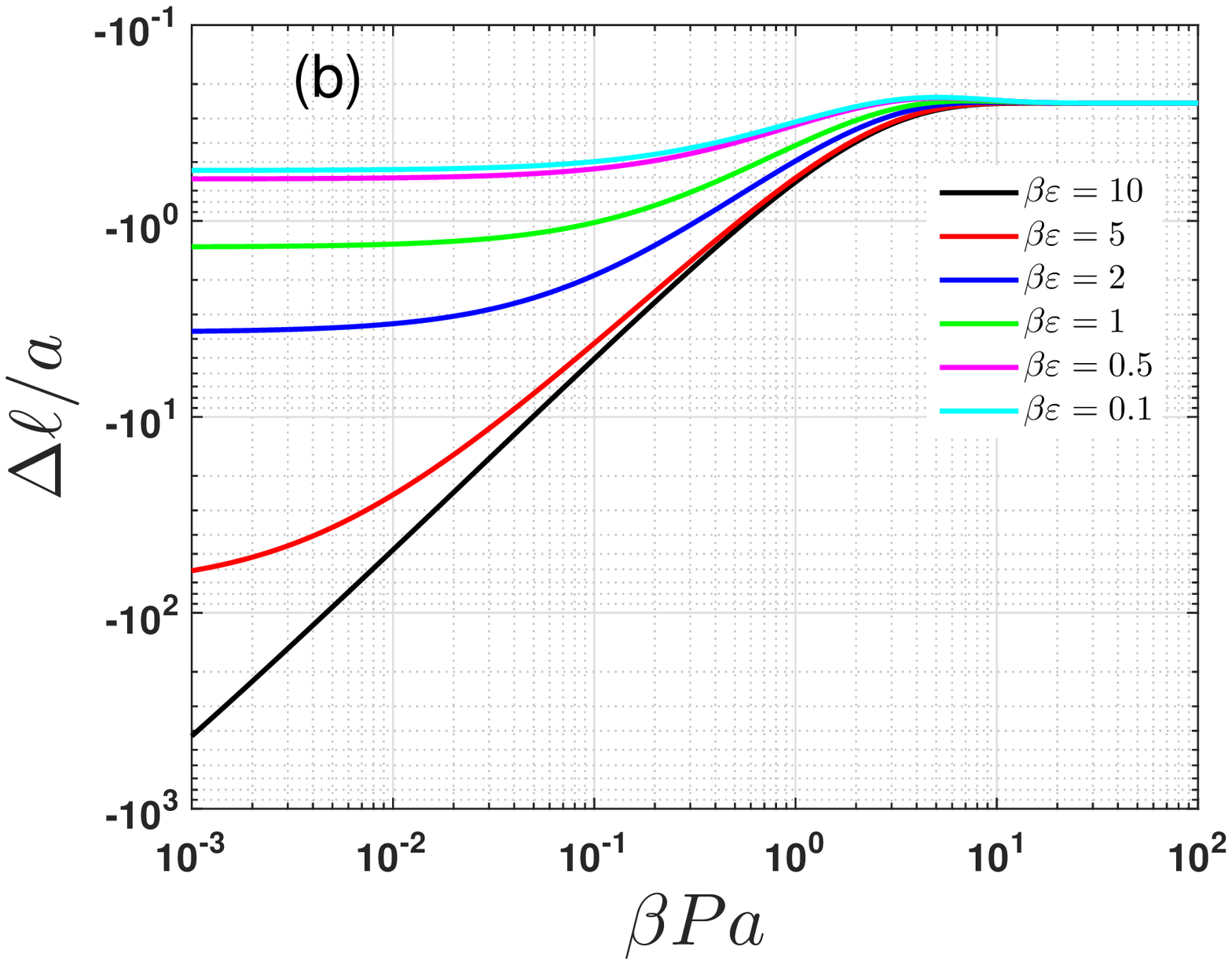}
	\caption{\label{fig:Length} (Color online) (a) Length per atom (in units of the bond range $a$) as a function of pressure (in units of $\varepsilon/a$) at various inverse temperatures (in units of $\varepsilon/k^{}_{\mathrm{B}}$), for $d/a=0.5$. (b) Length-per-atom difference, $\Delta\ell(P,T)=\ell(P,T)-\ell^{}_{\mathrm{HS}}(P,T)$, relative to hard spheres of diameter $a$, as a function of $\beta Pa$.}
\end{figure}

The length per atom $\ell(P,T)$ is plotted in Fig.~\ref{fig:Length}(a) as a function of pressure at various temperatures, for $d/a=0.5$. The various limits discussed above are more easily observed in a plot of the length difference, $\Delta\ell(P,T)=\ell(P,T)-\ell^{}_{\mathrm{HS}}(P,T)$ as a function of $\beta Pa$. Figure~\ref{fig:Length}(b) shows such a plot for the same data as in Fig.~\ref{fig:Length}(a). For $\beta P\left(a-d\right)\gg 1$ all curves approach the value $\Delta\ell/a=\left(d/a-1\right)/2=-0.25$ independent of the value of $\beta\varepsilon$, in agreement with Eq.~(\ref{eq:High_density_limit}). For $e^{-\beta\varepsilon}\ll\beta P\left(a-d\right)\ll 1$ Eq.~(\ref{eq:low_Temperature_density_limit}) gives $\Delta\ell/a\approx-\left(2\beta Pa\right)^{-1}$ which appears as a straight line on a log-log scale. This behavior is demonstrated by the black curve in Fig.~\ref{fig:Length}(b). Finally, in the limit $\beta P(a-d)\ll e^{-\beta\varepsilon}$ the correction to Eq.~(\ref{eq:High_Temperature_limit}) of order $\beta\varepsilon$ gives $\Delta\ell/a\approx-e^{\beta\varepsilon}\left(1-d/a\right)$. This is the value approached by the curves in Fig.~\ref{fig:Length}(b) for $\beta P(a-d)\ll e^{-\beta\varepsilon}$.

\begin{figure}[ht]
	\centering
	\includegraphics[width=0.52\textwidth,height=0.28\textheight]{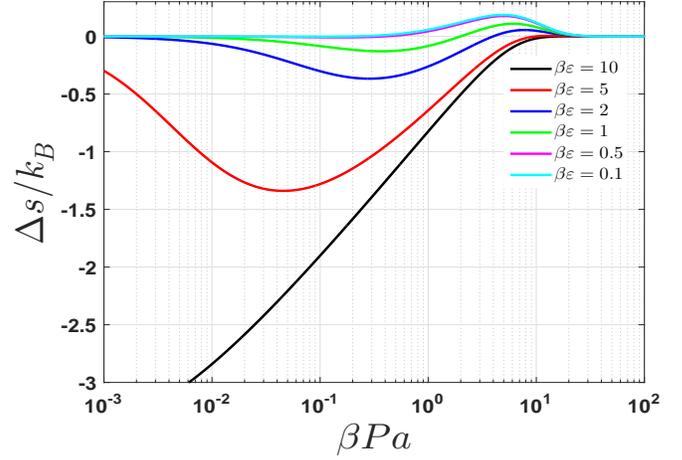}
	\caption{\label{fig:Entropy} (Color online) Entropy difference per atom, $\Delta s(P,T)=s(P,T)-s^{}_{\mathrm{HS}}(P,T)$, as a function of $\beta Pa$ at various temperatures, for $d/a=0.5$.}
\end{figure}

Figure~\ref{fig:Entropy} shows the entropy difference per atom, $\Delta s(P,T)=s(P,T)-s^{}_{\mathrm{HS}}(P,T)$, as a function of $\beta Pa$. The entropy difference approaches zero in the limits $\beta P\left(a-d\right)\gg 1$ or $\beta P(a-d)\ll e^{-\beta\varepsilon}$, in accordance with Eqs.~(\ref{eq:High_density_limit}) and~(\ref{eq:High_Temperature_limit}). It is also interesting to note that for $P=\left(a-d\right)/\varepsilon$ one has $A/B=e^{\beta\varepsilon}\left(e^{\beta\varepsilon}-1\right)$, and $\Delta g=g-g^{}_{\mathrm{HS}}$ [Eq.~(\ref{eq:Gibbs_bond})] becomes independent of temperature. Thus $\Delta s=0$ for $P=\left(a-d\right)/\varepsilon$, irrespective of the temperature.

\begin{figure}[ht]
	\centering
	\includegraphics[width=0.53\textwidth,height=0.28\textheight]{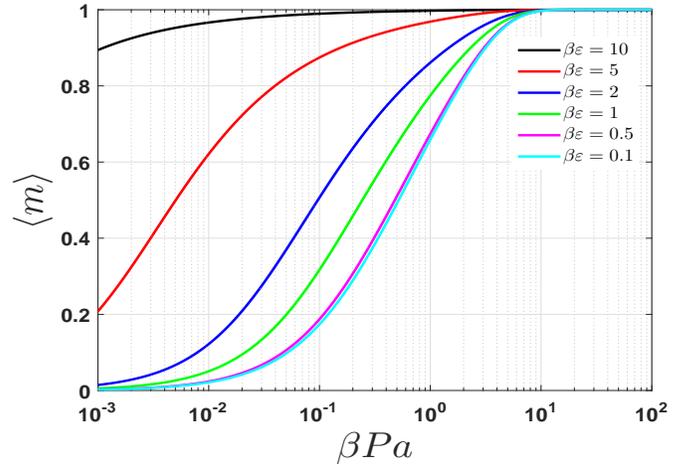}
	\caption{\label{fig:Fraction_molecules} (Color online) Fraction of molecules as a function of $\beta Pa$ at various temperatures, for $d/a=0.5$.}
\end{figure}

Another quantity of interest is the fraction of molecules
\begin{align}
\label{eq:Fraction_molecules}
\braket{m}&=A\lim^{}_{N\rightarrow\infty}\frac{1}{N/2}\frac{\partial\ln Z^{}_{N}(P,T)}{\partial A}=2A\frac{\partial\ln\Lambda^{}_{+}}{\partial A}\nonumber\\
&=\frac{\frac{4A}{B}}{\left(1+\sqrt{1+\frac{4A}{B}}\right)\sqrt{1+\frac{4A}{B}}}.
\end{align}
This quantity approaches 1 in the limit $\beta P\left(a-d\right)\gg e^{-\beta\varepsilon}$, and approaches 0 for $\beta P\left(a-d\right)\ll e^{-\beta\varepsilon}$, as shown in Fig.~\ref{fig:Fraction_molecules}. The first case contains the two limits $\beta P\left(a-d\right)\gg 1$ or $e^{-\beta\varepsilon}\ll\beta P\left(a-d\right)\ll 1$ discussed in Eqs.~(\ref{eq:High_density_limit}) and~(\ref{eq:low_Temperature_density_limit}), in which the system consists of $N/2$ molecules.

\subsubsection{The structure factor}

The transfer matrix method also enables the calculation of the structure factor of the system,
\begin{align}
\label{eq:Structure_factor1}
S(q)&=\frac{1}{N}\sum^{N}_{i,j=1}\braket{e^{-iq(x^{}_{i}-x^{}_{j})}}\nonumber\\
&=\frac{1}{N}\left(N+\sum^{N}_{i=1}\sum^{i-1}_{j=1}\braket{e^{-iq(x^{}_{i}-x^{}_{j})}}+\mathrm{c.c.}\right).
\end{align}
The inter-atomic distance can be written as a sum over bond lengths, $x^{}_{i}-x^{}_{j}=\sum^{i}_{k=j+1}y^{}_{k}$, and therefore one can factor the oscillatory term here, and use a variant of the transfer matrix method to calculate this quantity.  Adding the appropriate oscillatory factor to the expressions for $A$ and $B$ results in:
\begin{align}
\label{eq:a_and_b}
&A^{}_{q}=\int^{a}_{d}dye^{-iqy}e^{-\beta\left[v^{}_{\mathrm{bond}}(y)+Py\right]},\nonumber\\
&B^{}_{q}=\int^{\infty}_{a}dye^{-iqy}e^{-\beta\left[v^{}_{\mathrm{non-bond}}(y)+Py\right]},
\end{align}
together with an analogue of the transfer matrix:
\begin{align}
\label{eq:Transfer_matrix2}
&t=\begin{pmatrix}
B^{}_{q} & B^{}_{q} \\
A^{}_{q} & 0
\end{pmatrix}.
\end{align}
In terms of these quantities, one obtains
\begin{align}
\label{eq:Structure_factor2}
\braket{e^{-iq(x^{}_{i}-x^{}_{j})}}&=\frac{\mathrm{Tr}\left[T^{j}t^{i-j}T^{N-i}\right]}{\mathrm{Tr}\left[T^{N}\right]}\nonumber\\
&=\frac{\mathrm{Tr}\left[T^{N+j-i}t^{i-j}\right]}{\mathrm{Tr}\left[T^{N}\right]}.
\end{align}  

The eigenvalues of the matrix $t$ are $\lambda^{}_{\pm}=\left(B^{}_{q}\pm\sqrt{B^{2}_{q}+4A^{}_{q}B^{}_{q}}\right)/2$. Substituting the similarity transformations $T=MWM^{-1}$ and $t=mwm^{-1}$, where
\begin{align}
\label{eq:Matrices}
&W=\begin{pmatrix}
\Lambda^{}_{+} & 0 \\
0 & \Lambda^{}_{-}
\end{pmatrix},\nonumber\\
&w=\begin{pmatrix}
\lambda^{}_{+} & 0 \\
0 & \lambda^{}_{-}
\end{pmatrix},\nonumber\\
&M=\begin{pmatrix}
1 & 1 \\
\frac{A}{\Lambda^{}_{+}} & \frac{A}{\Lambda^{}_{-}}
\end{pmatrix},\nonumber\\
&m=\begin{pmatrix}
1 & 1 \\
\frac{A^{}_{q}}{\lambda^{}_{+}} & \frac{A^{}_{q}}{\lambda^{}_{-}}
\end{pmatrix},
\end{align}
we obtain from Eq.~(\ref{eq:Structure_factor2})
\begin{align}
\label{eq:Structure_factor3}
\braket{e^{-iq(x^{}_{i}-x^{}_{j})}}&=\frac{\mathrm{Tr}\left[MW^{N+j-i}M^{-1}mw^{i-j}m^{-1}\right]}{\mathrm{Tr}\left[T^{N}\right]}\nonumber\\
&=\frac{\mathrm{Tr}\left[m^{-1}MW^{N+j-i}M^{-1}mw^{i-j}\right]}{\mathrm{Tr}\left[T^{N}\right]}.
\end{align}
In the thermodynamic limit $N\rightarrow\infty$ only the mode with eigenvalue $\Lambda^{}_{+}$ survives, and one ends up with
\begin{align}
\label{eq:Structure_factor4}
\braket{e^{-iq(x^{}_{i}-x^{}_{j})}}=r^{}_{+}\left(\frac{\lambda^{}_{+}}{\Lambda^{}_{+}}\right)^{i-j}+r^{}_{-}\left(\frac{\lambda^{}_{-}}{\Lambda^{}_{+}}\right)^{i-j},
\end{align}
where
\begin{align}
\label{eq:Structure_factor5}
r^{}_{+}&=\left(m^{-1}M\right)^{}_{11}\left(M^{-1}m\right)^{}_{11}\nonumber\\
&=\frac{\Lambda^{}_{+}\lambda^{}_{+}+\Lambda^{}_{-}\lambda^{}_{-}-\Lambda^{}_{+}\Lambda^{}_{-}A^{}_{q}/A-\lambda^{}_{+}\lambda^{}_{-}A/A^{}_{q}}{\left(\Lambda^{}_{+}-\Lambda^{}_{-}\right)\left(\lambda^{}_{+}-\lambda^{}_{-}\right)},\nonumber\\
r^{}_{-}&=\left(m^{-1}M\right)^{}_{21}\left(M^{-1}m\right)^{}_{12}\nonumber\\
&=\frac{\Lambda^{}_{+}\Lambda^{}_{-}A^{}_{q}/A+\lambda^{}_{+}\lambda^{}_{-}A/A^{}_{q}-\Lambda^{}_{+}\lambda^{}_{-}-\Lambda^{}_{-}\lambda^{}_{+}}{\left(\Lambda^{}_{+}-\Lambda^{}_{-}\right)\left(\lambda^{}_{+}-\lambda^{}_{-}\right)}.
\end{align}
With the form~(\ref{eq:Structure_factor5}), the limit $N\rightarrow\infty$ of Eq.~(\ref{eq:Structure_factor1}) gives
\begin{align}
\label{eq:Structure_factor6}
S(q)=1+2\Re\left[\frac{r^{}_{+}\lambda^{}_{+}/\Lambda^{}_{+}}{1-\lambda^{}_{+}/\Lambda^{}_{+}}\right]+2\Re\left[\frac{r^{}_{-}\lambda^{}_{-}/\Lambda^{}_{+}}{1-\lambda^{}_{-}/\Lambda^{}_{+}}\right].
\end{align}
The corresponding radial distribution function $g(x)$ is given by the Fourier transform
\begin{align}
\label{eq:Radial_distribution_function}
g(x)=\frac{1}{2\pi n}\int dqe^{iqx}\left(S(q)-1\right),
\end{align}
where $n=\ell^{-1}$ is the particle density. 

\begin{figure}[ht]
	\centering
	\includegraphics[width=0.52\textwidth,height=0.28\textheight]{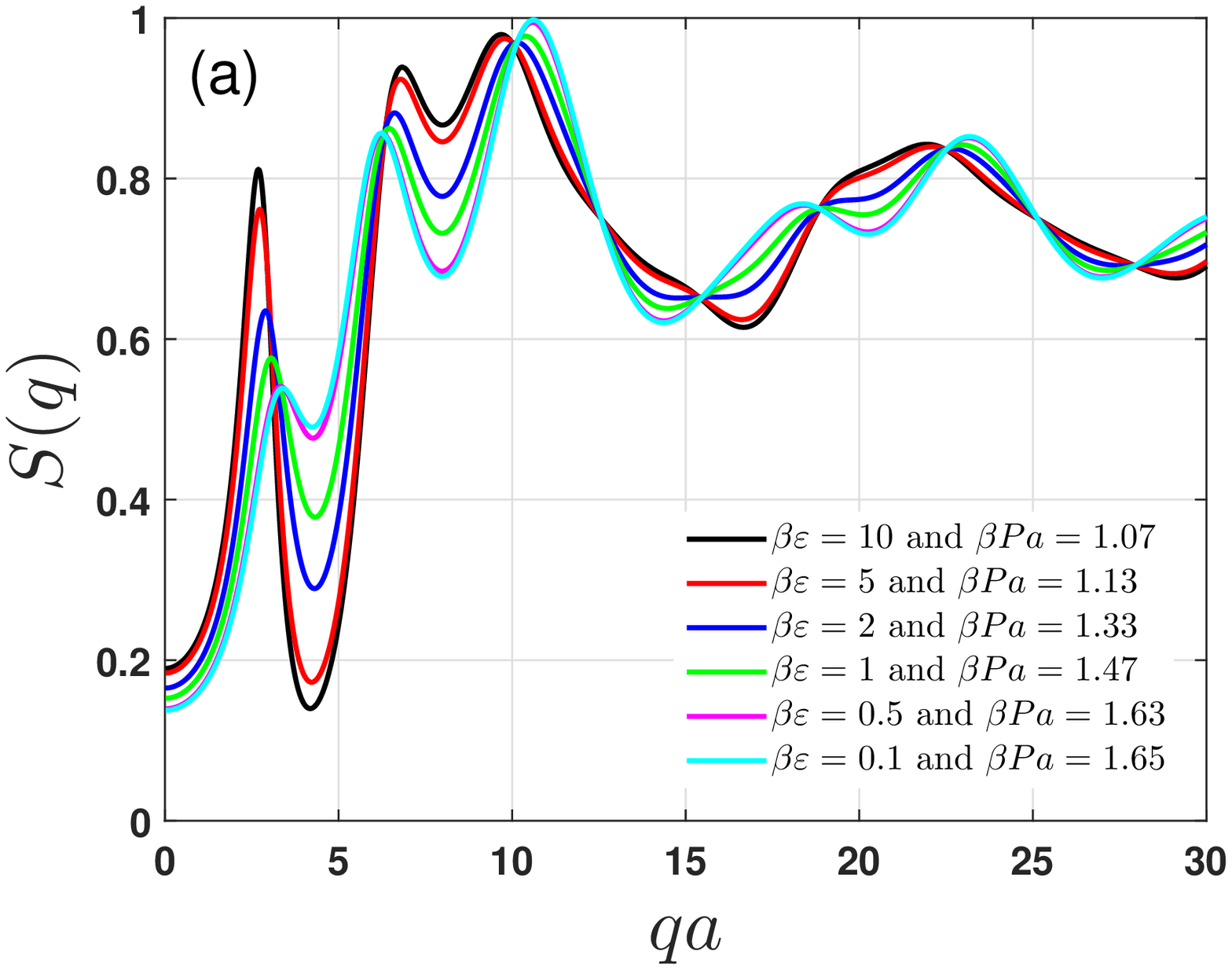}
	\includegraphics[width=0.52\textwidth,height=0.28\textheight]{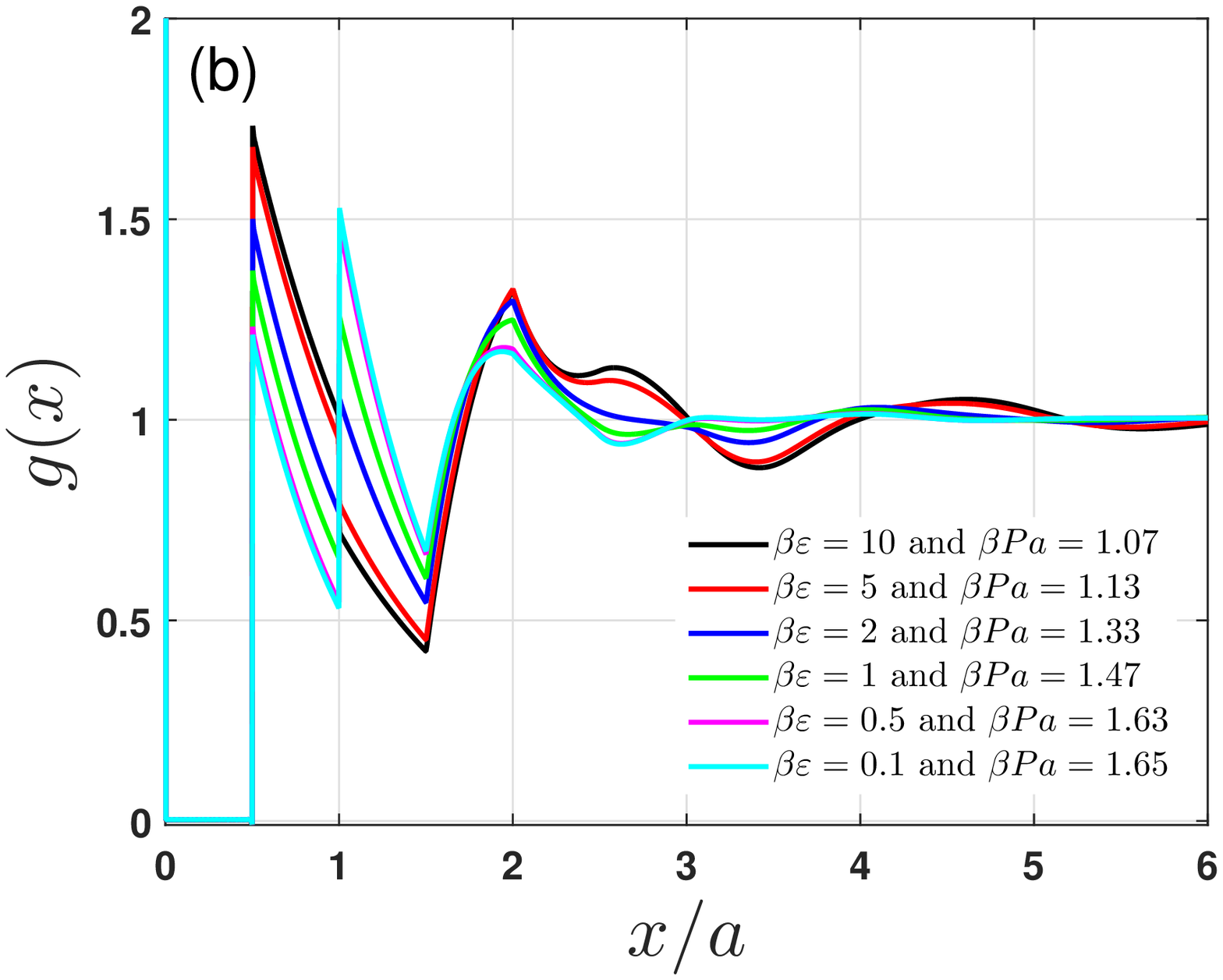}
	\caption{\label{fig:Structure} (Color online) (a) Structure factor $S(q)$ ($q$ in units of $a^{-1}$) at a constant density $na=0.75$, at various pressures and temperatures. (b) The corresponding radial distribution function $g(x)$ ($x$ in units of $a$).}
\end{figure}

The structure factor and radial distribution function for the case $d/a=0.5$, 
$v^{}_{\mathrm{non-bond}}(x)=0$ and $v^{}_{\mathrm{bond}}(x)=-\varepsilon$ discussed above are shown in Fig.~\ref{fig:Structure}, for a constant density $na=0.75$ (note that at a constant density the values of $\beta Pa$, which are given in the figure, vary only to a limited extent; a plot of the structure factors at a fixed value $\beta Pa\approx 1$ will be qualitatively similar). At low temperatures the structure factor shown in the upper panel has a pronounced first peak, which decreases in height and develops into a shoulder-like feature as the temperature is increased. The radial distribution function displayed in the lower panel has two main features corresponding to neighbors at distance $x=d$ and $x=a$, with discontinuities reflecting the discontinuities of the potential. At high temperatures these two features are distinct peaks, and the height of the second one is larger, since molecules are dissociated into individual atoms. At lower temperatures, the number of bonded molecules increases and the height of the first peak becomes dominant. Interestingly, at low enough temperatures the discontinuity at $x=a$ changes sign, and there is no longer a peak at this value of $x$. The evolution of this discontinuity with temperature is reflected in the strong oscillations observed in the upper panel for large $q$.

\subsubsection{Phases and boundaries}

The above, especially Fig.~\ref{fig:Fraction_molecules}, shows how this system transforms continuously between a purely atomic fluid and a molecular fluid. It is intuitively clear that a qualitatively similar crossover will occur also in 2D and 3D. In particular, in this 1D model the $P \to 0$ and the $T \to 0$ limits are different and do not commute, and there is in this sense a discontinuity at $T=P=0$. A similar discontinuity has been identified in previous studies of 1D systems as representing the liquid-gas critical point of 3D fluids (see, e.g., Sec.~VII.A of Ref.~\onlinecite{SLMR99}). It is worth noting that for a system defined just as above but with a negative value of $\varepsilon$, this critical point occurs at a finite pressure, $P^{}_{\mathrm{C}}=|\varepsilon|/(a-d)$, as follows from Eq.~(\ref{eq:Fraction_molecules}). In this case, at low pressure the atoms repel each other and are separated by distances larger than $a$, but a pressure $P>P^{}_{\mathrm{C}}$ causes pairs of atoms to overcome their mutual repulsion. This transition is similar to that found in a system of particles interacting through soft-core pair potentials, see Refs.~\onlinecite{BSV09} and~\onlinecite{SLMR99}, and references therein.

Returning to the present system with a positive binding energy, one can identify additional features related to phase behavior. If the temperature is low enough, the fraction of molecules approaches 100\%, and only very few unpaired atoms remain. The system then consists of large domains of molecules, separated by ``boundary defects'' which are unbound atoms, and which can only be created or annihilated in pairs. In this sense, the single atoms act as topological defects in the system.

The dynamics of these defects may be of interest. If the temperature or pressure is slightly changed, the density of defects will adjust and approach the new equilibrium very slowly, due to the need to break the strong covalent bonds for any change to occur. After the kinetic energy of all the atoms has relaxed to the new temperature, changes in the position of a defect along the chain require $\propto\exp(\beta\varepsilon)$ attempts, because of the large barrier for breaking these bonds. Such bond-breaking events enable just a single step along the chain, but in order to annihilate two defects must approach each other by diffusing significant distances along the chain.

Conversely, it is not enough to break a single molecular bond in order to create a pair of defects, because there is an overwhelming probability for the two single atoms just created to recombine, leaving the system in the original state. Two adjacent molecules in a given domain must simultaneously have broken bonds in order for a single novel molecule to form, nucleating a new domain, and creating two boundary defects. The probability of this happening is $\propto \exp(-2\beta\varepsilon)$ per attempt, and it must further be taken into account that once the two newborn defects begin diffusing, the probability that they will recombine is relatively large, because of their proximity to each other. Only after a long relaxation process will the concentration of defects reach its new equilibrium value, and the creation and annihilation processes reach equal rates (detailed balance).

\subsection{Bond-counting potential with $\mathbf{z=2}$ and double bonds} \label{z=2}

We now extend the treatment of the previous subsection to a model with two bonds per atom, where each atom can have up to two single bonds (one on each side), or one double bond with one of its nearest neighbors. We assume that $d>a/2$, so that next-nearest neighbors cannot penetrate the bonding zone and two single bonds on the same side are forbidden.~\cite{Comment2} Each single bond has an energy $-\varepsilon$ with bond length $\tilde{a}<x<a$, and a double bond contributes an energy $-\tilde{\varepsilon}$ with bond length $d<x<\tilde{a}$. The appropriate transfer matrix is
\begin{align}
\label{eq:Transfer_matrix3}
&T=\begin{pmatrix}
	C & C & C \\
	B & B & 0 \\
	A & 0 & 0
\end{pmatrix},
\end{align}
where $A$, $B$, and $C$ are defined in analogy with Eqs.~(\ref{eq:A}) and~(\ref{eq:B}). Powers of this matrix contain terms where double bonds, $A$, are always flanked by non-bonds, $C$, whereas single bonds, $B$, are not limited in this manner (but cannot, of course, be adjacent to a double bond, $A$). The phase in which each pair of atoms has a double bond (i.e., the state whose contribution to the partition function is $\ldots ACACAC\ldots$) corresponds to the molecular phase, whereas the phase in which each atom is bonded to its two neighbors by single bonds (i.e., the state whose contribution to the partition function is $\ldots BBBBBB\ldots$) corresponds to a polymeric liquid. 

Consider a possible transition between these two phases at $T=0$. The enthalpies associated with each of these bonds are $H^{}_{A}(P)=-\tilde{\varepsilon}+Pd$, $H^{}_{B}(P)=-\varepsilon+P\tilde{a}$ and $H^{}_{C}(P)=Pa$ (this is the minimum enthalpy obtained for the minimum bond length in each case). The corresponding enthalpies per atom are
\begin{align}
\label{eq:Enthalpies}
&H^{}_{\mathrm{mol}}(P)=\frac{H^{}_{A}(P)+H^{}_{C}(P)}{2}=-\frac{\tilde{\varepsilon}}{2}+P\cdot\frac{\left(d+a\right)}{2},\nonumber\\
&H^{}_{\mathrm{pol}}(P)=H^{}_{B}(P)=-\varepsilon+P\tilde{a}.
\end{align}  
A transition will occur for the case $\tilde{\varepsilon}>2\varepsilon$ and $\left(d+a\right)>2\tilde{a}$, for which the molecular phase is energetically favorable but the polymeric phase becomes favorable as the pressure is increased. The enthalpies of the two phases are equal at the critical pressure
\begin{align}
\label{eq:Critical_pressure}
&P^{}_{\mathrm{C}}=\frac{\tilde{\varepsilon}-2\varepsilon}{d+a-2\tilde{a}}.
\end{align}

This model has sufficient detail to have a finite free-energy cost for a domain boundary. In order to calculate this cost, consider replacing a double bond by a single one, namely, replacing one of the $A$'s in the term $\ldots ACACAC\ldots$ by $B$. Introducing a larger domain of $B$'s  at $T=0$ and $P=P^{}_{\mathrm{C}}$ will cost the same free energy, 
\begin{align}
\label{eq:DeltaH}
\Delta H & =H^{}_{B}(P^{}_{\mathrm{C}})-H^{}_{A}(P^{}_{\mathrm{C}}) 
\nonumber\\
& = -\varepsilon+\tilde{\varepsilon}+P^{}_{\mathrm{C}}\left(\tilde{a}-d\right)
\nonumber\\
& = \varepsilon+\left(\tilde{\varepsilon}-2\varepsilon\right)
\frac{a-\tilde{a}}{d+a-2\tilde{a}}
\nonumber\\
& >\varepsilon .
\end{align}
Thus, the two phases are not mixed --- for any length per atom in the range $\tilde{a}<\ell<(d+a)/2$, the system will have a clear tendency to phase-separate at zero temperature.

As per Landau's argument for the absence of phase transitions in 1D systems, entropic effects will always be dominant in the thermodynamic limit, and the resulting finite concentration of phase-boundary defects will round off the phase transformation. However, as the energy scales for covalent bonds can easily be more than an order of magnitude larger than room temperature, it makes sense to also consider the behavior of finite systems; for a system of $N$ atoms, the whole system will be predominantly in one phase for $k^{}_{\mathrm{B}}T\ll\Delta H / \ln N$ (in principle, at finite temperatures $\Delta H$ should be replaced by $\Delta G$).

This transition between a molecular phase and a polymeric phase, obtained for a 1D model system, is closely analogous to the known LLPT in phosphorus discussed in the introduction.~\cite{KY00,KY02,MG03} Due to the phase-separation just discussed, the signature of a first-order phase transition is clearer in this case, when compared to the signatures of phase transitions in Takahashi solution to simpler 1D models with nearest-neighbor pair potentials.~\cite{SLMR99} The reason for this is that in such solutions, the type of bonding between two neighboring atoms has no effect on their other neighbors, and thus no mechanism for forming a finite domain-boundary energy exists.

A variant of the present model can also reproduce a non-equilibrium phase transition, closely analogous to boiling of white phosphorus. In order to obtain a boiling transition, one must include weak attractive forces between molecules (in $v^{}_{\mathrm{non-bond}}$). The polymerization transition can be dynamically hindered at low temperatures, by preparing the system as a low density molecular gas, and then bringing about condensation, i.e., increasing the density in a non-equilibrium manner. If the binding energy of each doubly-bonded molecule is much larger than the temperature (say, by a factor of 50), one can reach the condensed molecular phase, even if the polymeric phase is the thermodynamically stable one, because nucleation events of the polymeric phase will be exceedingly rare. Next, one may fix pressure and change the temperature. In 3D, one may have different solid, liquid and gaseous molecular phases, but for molecules in 1D, only a fluid phase exists. The transition between a liquid-like and a gas-like system is gradual in 1D, rather than abrupt, but one may still define a characteristic temperature for this transition. If this temperature is no longer very small compared to the molecular binding energy, nucleation of the polymeric phase is expected to occur at a rate which is no longer negligible. This is qualitatively what happens in actual experiments on boiling of white phosphorus.~\cite{DTW46}

\section{Summary and discussion} \label{Discussion}

We have suggested a simple type of classical many-body model potential, which imposes constraints on (i.e., counts) the number of bonds allowed for each atom, thus describing the essence of chemical valency. An analytical solution of the equilibrium statistical mechanics of a one-bond-per-atom model ($z=1$) in 1D using the transfer matrix method, including the relative density of diatomic molecules and the structure factor, was presented. Our solution generalizes the Takahashi solution for nearest-neighbor pair potentials. The results for the specific case of a constant attractive potential $-\varepsilon$ ($\varepsilon>0$) inside the bonding zone ($d<x<a$, where $d$ and $a$ are the hard core and bonding zone diameters) and zero outside of it ($x>a$) were displayed in detail.

The case of $z=2$ with double bonds was also discussed. This model exhibits a first-order phase transition (at $T=0$ in the thermodynamic limit) between a molecular phase and a polymeric phase, reminiscent of the LLPT observed in phosphorus. Furthermore, this model can also describe a non-equilibrium phase transformation analogous to the boiling of white phosphorus.

There are many open directions for future work with bond-counting potentials. One may introduce greater variety, by considering different valencies $z$ and also mixtures of several different species of atoms. One may study more realistic separation-dependent potentials rather than constants, or more complicated $q_i,q_j$-dependence of $\mathcal{V}^{}_{ij}$ (see Sec.~\ref{Model}). In particular, the potential felt by a single atom approaching another single atom could be attractive and continuous at $y=a$, while at the same time the potential for an atom approaching a molecule could be repulsive already for $y>a$. On the other hand, one may study even simpler systems by considering, e.g., the case $d=\epsilon=0$ (or $\epsilon \to \infty$), resulting in a system which has only the integer $z$ parameter in addition to the length parameter $a$. As for the hard-sphere model, the temperature-dependece in such a system is trivial, and the properties, including phase transitions, depend only on $z$ and the reduced density $na^{3}$.

An extension of the model potentials proposed in this paper to 2D and 3D using molecular dynamics simulations may help to gain further insight into the mechanisms responsible for LLPTs and liquid state anomalies. A particularly interesting question is the mechanism behind the LLPTs in bismuth. In fact, one of the present authors has speculated that the covalently-bonded bilayer sheets of crystalline bismuth become only partly disordered at melting ($545\,$K), and fully disintegrate only at the LLPT at $1010\,$K.~\cite{AN10} This would mean that the phase obtained by melting of solid bismuth consists of 2D polymers (the bilayers), which are unable to pass through each other, but able to glide in the transverse directions, possibly becoming crumpled, thus adding to the disorder.

This suggested mechanism also provides a reasonable context for explanation of the recently-observed anomalies of Refs.~\onlinecite{SY17,SY16a,SY16b}, concerning bismuth samples which underwent a high-pressure and high-temperature treatment. The 2D polymers which could result form the cooling and solidification procedure at $2\,$GPa, a pressure at which there is hardly any difference in density between the liquid and the solid, would perhaps have nontrivial topologies --- possibly the sheets are rolled up into tubes which are entangled on larger scales. This assumes that the covalent bonds are strong enough to remain intact during the phase transitions from the melt to the ``bismuth II'' and then to the ``bismuth I'' solid phases (the quotes are a reminder that these samples contain significant structural defects;~\cite{SY16a} the character of these defects is consistent with this type of explanation). The stability of the complex topology of these sheets could then also be responsible for the absence of flow in the ``melted'' samples.~\cite{SY16b} An encouraging hint comes from additional studies of the phase behavior of bismuth at $\sim 2\,$GPa and elevated temperatures, which may be interpreted as indicating that at these pressures and at temperatures near melting the bonds retain a certain degree of stability even while the sample undergoes a phase transition. These studies identified cases in which the type of phase observed in this range of $P$ and $T$ depended on the path previously followed, i.e., on which phase the sample was previously in.~\cite{LC17,PE06}

It is natural to study the qualitative aspects of such suggested mechanisms using a $z=3$ bond-counting potential in 3D. The complex topology of crumpled 2D polymers suggests that topological defects will play an important role in molecular dynamics or Monte Carlo simulations of such systems. In this sense, the present 1D work may be viewed as a preliminary step in developing this approach, involving topological defects which are particularly simple.

\begin{acknowledgments}
We acknowledge fruitful discussions with E. Pazy and V. V. Brazhkin.
\end{acknowledgments}


\begin{references}
	\bibitem{Hansen&McDonald} J.-P. Hansen and I. R. McDonald, {\it Theory of Simple Liquids: with Applications to Soft Matter}, 4th edn (New York, Academic, 2013).
	\bibitem{CD83} D. Chandler, J. D. Weeks, and H. C. Andersen {\bf 220}, 787 (1983).
	\bibitem{ANW66} N. W. Ashcroft and J. Lekner, Phys. Rev. {\bf 145}, 83 (1966).
	\bibitem{YJL73} J. L. Yarnell, M. J. Katz, R. G. Wenzel, and S. H. Koenig, Phys. Rev. A {\bf 7}, 2130 (1973).
	\bibitem{DJC16} J. C. Dyre, J. Phys.: Condens. Matter {\bf 28}, 323001 (2016).
	\bibitem{LJE24} J. E. Lennard-Jones, Proc. R. Soc. A {\bf 106}, 441 (1924).
	\bibitem{BVV97} V. V. Brazhkin, S. V. Popova, and R. N. Voloshin, High Press. Res. {\bf 15}, 267 (1997).
	\bibitem{BVV02} V. V. Brazhkin, S. V. Buldyrev, V. N. Ryzhov, and H. E. Stanley, {\it New Kinds of Phase transitions: Transformations in Disordered Substances} (Kluwer, Dordrecht, 2002).
	\bibitem{BVV03} V. V. Brazhkin and A. G. Lyapin, J. Phys.: Condens. Matter {\bf 15}, 6059 (2003).
	\bibitem{MPF07} P. F McMillan, M. Wilson, M. C Wilding, D. Daisenberger, M. Mezouar, and G N. Greaves, J. Phys.: Condens. Matter {\bf 19}, 415101 (2007).
	\bibitem{KY00} Y. Katayama, T. Mizutani, W. Utsumi, O. Shimomura, M. Yamakata, and K. Funakoshi, Nature {\bf 403}, 170 (2000).
	\bibitem{KY02} Y. Katayama, J. Non-Cryst. Solids {\bf 312-314}, 8 (2002).
	\bibitem{MG03} G. Monaco, S. Falconi, W. A. Crichton, and M. Mezouar, Phys. Rev. Lett. {\bf 90}, 255701 (2003).
	\bibitem{BVV89} V. V. Brazhkin, S. V. Popova, and R. N. Voloshin, JETP Lett {\bf 50}, 362 (1989).
	\bibitem{BVV91} V. V. Brazhkin, R. N. Voloshin, S. V. Popova, and A. G. Umnov, Phys. Lett. A {\bf 154}, 413 (1991).
	\bibitem{LL14} L. Liu, Y. Kono, C. Kenney-Benson, W. Yang, Y. Bi and G. Shen, Phys. Rev. B {\bf 89}, 174201 (2014).
	\bibitem{ZG14} G. Zhao and H. Mu, Chem. Phys. Lett. {\bf 616-617}, 131 (2014).
	\bibitem{HL18} L. Henry, M. Mezouar, G. Garbarino, D. Sifr\'{e}, G. Weck, and F. Datchi, arXiv:1709.09996.
	\bibitem{UAG92} A. G. Umnov, V. V. Brazhkin, S. V. Popova, and R. N. Voloshin, J. Phys: Condens. Matter {\bf 4}, 1427 (1992).
	\bibitem{GY09} Y. Greenberg, E. Yahel, E. N. Caspi, C. Benmore, B. Beuneu, M. P. Dariel and G. Makov, Europhys. Lett. {\bf 86}, 36004 (2009).
	\bibitem{SY17} Y. Shu, D. Yu, W. Hu, Y. Wangb, G. Shen, Y. Kono, B. Xu, J. He, Z. Liu, and Y. Tian, PNAS {\bf 114}, 3375 (2017).
	\bibitem{EM18} M. Emuna, S. Matityahu, E. Yahel, G. Makov, and Y. Greenberg, J. Chem. Phys. {\bf 148}, 034505 (2018).
	\bibitem{TY85} Y. Tsuchiya and E. F. W. Seymour, J. Phys. C: Solid State Phys. {\bf 18}, 4721 (1985).
	\bibitem{BVV92} V. V. Braznkin, R. N. Voloshin, S. V. Popova, and A. G. Umnov, J. Phys.: Condens. Matter {\bf 4}, 1419 (1992).
	\bibitem{BVV06} V. V. Brazhkin, J. Phys.: Condens. Matter {\bf 18}, 9643 (2006).
	\bibitem{Phase Diagrams} E. Yu. Tonkov and E. G. Ponyatovsky, {\it Phase Transformations of Elements Under High Pressure} (CRC Press LLC, 2005).
	\bibitem{DTW46} T. W. DeWitt and S. Skolnik, J. Am. Chem. Soc. {\bf 68}, 2305 (1946).
	\bibitem{LXF07} X.-F. Li, F.-Q Zu, L.-J Liu, J. Yu, and B. Zhou, Phys. Chem. Liq. {\bf 45}, 531 (2007).
	\bibitem{SY16a} Y. Shu, W. Hu, Z. Liu, G. Shen, B. Xu, Z. Zhao, J. He, Y. Wang, Y. Tian, and D. Yu, Sci. Rep. {\bf 6}, 20337 (2016).
	\bibitem{SY16b} Y. Shu, W. Hu, Z. Zhao, L. Wang, Z. Liu, Y. Tian, and D. Yu, Mater. Lett. {\bf 168}, 36 (2016).
	\bibitem{SS65} S. Str\"{a}ssler, C. Kittel, Phys. Rev. {\bf 139}, A758 (1965).
	\bibitem{RE67} E. Rapoport, J. Chem. Phys. {\bf 46}, 2891 (1967).
	\bibitem{TH00} H. Tanaka, Phys. Rev. E {\bf 62}, 6968 (2000).
	\bibitem{PEG03}  E. G. Ponyatovsky, J. Phys.: Condens. Matter {\bf 15}, 6123 (2003).
	\bibitem{twostatevariants} Focussing on the energetic and volumetric properties of the two types of clusters, and omitting the entropic aspect, leads to a restricted parameter space, in which the model exhibits only positive slopes of the LLPTs, as in Ref. \onlinecite{BVV97}.
	\bibitem{MT01} T. Morishita, Phys. Rev. Lett. {\bf 87}, 105701 (2001).
	\bibitem{GM05} L. M. Ghiringhelli and E. J. Meijer, J. Chem. Phys. {\bf 122}, 184510 (2005).
	\bibitem{PHP03} P. H. Poole, F. Sciortino, U. Essmann, and H. E. Stanley, Nature (London) {\bf 360}, 324 (1992).
	\bibitem{SS03} S. Sastry and C. A. Angell, Nat. Mater. {\bf 2}, 739 (2003).
	\bibitem{WG17} G. Weck, F. Datchi, G. Garbarino, S. Ninet, J.-A Queyroux, T. Plisson, M. Mezouar, and P. Loubeyre, Phys. Rev. Lett {\bf 119}, 235701 (2017).
	\bibitem{ABJ57} B. J. Alder and T. E. Wainwright, J. Chem. Phys. {\bf 27}, 1208 (1957).
	\bibitem{WWW57} W. W. Wood and J. D. Jacobson, J. Chem. Phys. {\bf 27}, 1207 (1957).
	\bibitem{HWG68} W. G. Hoover and F. H. Ree, J. Chem. Phys. {\bf 49}, 3609 (1968).
	\bibitem{BSV09} S. V. Buldyrev, G. Malescio, C. A. Angell, N. Giovambattista, S. Prestipino, F. Saija, H. E. Stanley, and L. Xu, J. Phys.: Condens. Matter {\bf 21}, 504106 (2009).
	\bibitem{CCH96} C. H. Cho, S. Singh, and G. W. Robinson. Phys. Rev. Lett. {\bf 76}, 1651 (1996).
	\bibitem{SLMR99} M. R. Sadr-Lahijany, A. Scala, S. V. Buldyrev, and H. E. Stanely, Phys. Rev. E {\bf 60}, 6714 (1999).
	\bibitem{BNA08} A. Ben-Naim, J. Chem. Phys. {\bf 128}, 024505 (2008).
	\bibitem{DMS93} M. S. Daw, S. M. Foiles, and M. I. Baskes, Mater. Sci. Rep. {\bf 9}, 251 (1993).
	\bibitem{STP16} T. P. Senftle, S. Hong, M. M. Islam, S. B. Kylasa, Y. Zheng, Y. K. Shin, C. Junkermeier, R. Engel-Herbert, M. J. Janik, H. M. Aktulga, T. Verstraelen, A. Grama, and A. C. T. van Duin, npj Comput. Mater. {\bf 2}, 15011 (2016).
	\bibitem{WMS84} M. S. Wertheim, J. Stat. Phys. {\bf 35}, 19 (1984); M. S. Wertheim, J. Stat. Phys. {\bf 35}, 35 (1984).
	\bibitem{JG88} G. Jackson, W. G. Chapman, and K. E. Gubbins, Mol. Phys. {\bf 65}, 1 (1988).
	\bibitem{CWG} W. G. Chapman, K. E. Gubbins, G. Jackson, and M. Radosz, Fluid Phase Equil. {\bf 52}, 31 (1989); W. G. Chapman, K. E. Gubbins, G. Jackson, and M. Radosz, Ind. Eng. Chem. Res. {\bf 29}, 1709 (1990).
	\bibitem{Baskes92} M. I. Baskes, Phys. Rev. B {\bf 46}, 2727 (1992).
	\bibitem{TH42} H. Takahashi, Proc. Phys. Math. Soc. Jpn. {\bf 24}, 60 (1942).
	\bibitem{Lieb&Mattis} E. H. Lieb and D. C. Mattis, {\it Mathematical Physics in One Dimension} (Academic, New York, 1966).
	\bibitem{Comment1} Note that for particles with hard core of diameter $a$, one can state of the problem by means of a pair potential $U(x)$ and Hamiltonian
	$\mathcal{H}(\mathbf{x},\mathbf{p})=\frac{1}{2m}\sum^{N}_{i=1}p^{2}_{i}+\sum^{N}_{i=1}\sum^{i-1}_{j=1}U(|x^{}_{i}-x^{}_{j}|)$, provided that $U(x)=0$ for $x>2a$, because then the hard-sphere constraint implies that there is no interaction between next-nearest neighbors. However, the Hamiltonian~(\ref{eq:Hamiltonian1}) with nearest-neighbor interactions is more general, as it allows for $v(x)\neq 0$ for $x>2a$.
	\bibitem{BNA92} A. Ben-Naim, {\it Statistical Thermodynamics for Chemists and Biochemists} (New York, Plenum Press, 1992). 
	\bibitem{Comment2} Removing this restriction requires generalizing Eq.~(\ref{eq:A}) to include multiple integrals, using several configuration variables $y$ to describe the more-complicated structure of the resulting molecules.
	\bibitem{AN10} N. Argaman, Phys. Lett. A {\bf 374}, 3982 (2010).
	\bibitem{LC17} C. Lin, J. S. Smith, S. V. Sinogeikin, Y. Kono, C. Park, C. Kenney-Benson, and G. Shen, Nat. Commun. {\bf 8}, 14260 (2017).
	\bibitem{PE06} E. Principi, M. Minicucci, A. Di Cicco, A. Trapananti, S. De Panfilis, and R. Poloni, Phys. Rev. B {\bf 74}, 064101 (2006).
\end{references}
\end{document}